\documentclass[10pt,twocolumn]{article} 
\usepackage{simpleConference}
\usepackage{times}
\usepackage{graphicx}
\usepackage{subfig}
\usepackage{amssymb}
\usepackage{enumitem}
\usepackage{url,hyperref}
\usepackage{tikz}
\usepackage{footnote}
\usepackage{multirow}
\usetikzlibrary{shapes,arrows}

\makesavenoteenv{tabular}
\begin{document}
\title{Lesion Analysis and Diagnosis with Mask-RCNN}

\author{Andrey Sorokin \\
\\
Voronezh, Russia\\
\today
\\
\\
Andrey.I.Sorokin@gmail.com \\
}

\maketitle
\thispagestyle{empty}

\begin{abstract}
This project applies Mask R-CNN\cite{DBLP:journals/corr/HeGDG17} method to ISIC 2018 challenge tasks: lesion boundary segmentation (task 1), lesion attributes detection (task 2), lesion diagnosis (task 3). A proposed solution to the latter based on combining task 1 and 2 models and introducing a simple voting procedure.
\end{abstract}

\section{Baseline And Framework}
The results presented in this article were acquired using Mask R-CNN approach\cite{DBLP:journals/corr/HeGDG17} and a public implementation\footnote{\url{https://github.com/matterport/Mask_RCNN}} on Keras\footnote{\url{https://keras.io/}} and Tensorflow\footnote{\url{https://www.tensorflow.org/}} in Python 3.5. The source code and the setup details are publicly available in GIT repository\footnote{\url{https://bitbucket.org/rewintous/isic_2018_mrcnn_submit}}.
All experiments executed on a single NVIDIA Geforce 1080 with 8Gb of RAM.

\section{Training Data, Augmentation, Scoring}

ISIC 2018 Challenge dataset \cite{DBLP:dermato1/journals/corr/abs-1710-05006}\cite{DBLP:dermato2/journals/corr/abs-1803-10417} consists of the following data depending on the task.

\begin{enumerate}
	\item For boundary segmentation and attributes detection tasks:	
	\begin{flushleft}
	  	\begin{enumerate}
			\item[$\bullet$] 2,594 training lesion images;
			\item[$\bullet$] 2,594 binary mask images indicating lesion location;
			\item[$\bullet$] 12,970 binary mask images (5 per each training image) indicating the location of a dermoscopic attribute.
		\end{enumerate}
	\end{flushleft}
	\item For disease classification task:
	\begin{flushleft}
		\begin{enumerate}
			\item[$\bullet$] 10,015 training lesion images;
			\item[$\bullet$] 1 ground truth CSV file describing 10,015 ground truth diagnoses of the training dataset.
		\end{enumerate}
\end{flushleft}
\end{enumerate}

The input lesion images were split into training and test sets, containing correspondingly:

\begin{enumerate}
	\item[$\bullet$] 2,294 and 300 images for task 1 and 2;
	\item[$\bullet$] 8,015 and 2000 images for task 3.
\end{enumerate}

During training, images were augmented with:
\begin{enumerate}
	\item[$\bullet$] vertical and horizontal image flips;
	\item[$\bullet$] 90$^{\circ}$, 180$^{\circ}$, 270$^{\circ}$ image rotations;
	\item[$\bullet$] luminosity scaling within range: [0.8, 1.5];
	\item[$\bullet$] gaussian blur with standard deviation 2.5.
\end{enumerate}

The required format of tasks 1 and 2 prediction is one or more binary masks indicating either a lesion boundary or specific lesion attributes. A goal of tasks 1 and 2 is a maximization  of Jaccard index of predicted and ground truth masks:

\begin{equation}
 J(A,B) = {{|A \cap B|}\over{|A \cup B|}}, \quad 0\le J(A,B)\le 1,
 \label{eq:jccrd}
\end{equation}

\begin{flushleft}
	where A and B - sets of pixel coordinates, representing a ground truth and a predicted mask correspondingly.
\end{flushleft}

Task 3 prediction for an input image is a vector of diagnosis confidences for each disease. Task 3 predictions are validated using normalized multi-class accuracy metrics.

\section{Task 1: Lesion Boundary Segmentation} 
\subsection{Configuration}
The Mask-RCNN model was used for training with the following parameters:

\begin{tabular}{| l | l |}
    \hline
	Number of classes & 2 \\ \hline
	Backbone network & ResNet50\cite{DBLP:resnet/journals/corr/HeZRS15} \\ \hline	
	Input image dimensions & 768x768 \\ \hline
	RPN\footnote{RPN - Region Proposal Network, please, see \cite{DBLP:journals/corr/HeGDG17} for details} Anchor Scales & 
	32, 64, 128, 256, 512 \\ \hline
	Anchors per image & 64 \\ \hline
	Mask shape & 56x56 \\ \hline
	Train RoIs\footnote{RoI - Region of Interest, please, see \cite{DBLP:journals/corr/HeGDG17} for details} per image & 128 \\ \hline
	Learning Rate & 0.001 \\ \hline	
	Learning Momentum & 0.9 \\ \hline	
	Weight Decay & 0.0001 \\ \hline	
\end{tabular}

\hfill \\

Every image and mask from training set was resized to 768 pixels along the longest side and padded to match required input dimensions of 768 by 768 pixels. The input image size was chosen experimentally to fit the model in GPU memory.

\begin{figure}[b!]
	\centering
	\begin{tabular}{ccc}
		\subfloat[input image]{\includegraphics[width = 1.0in]{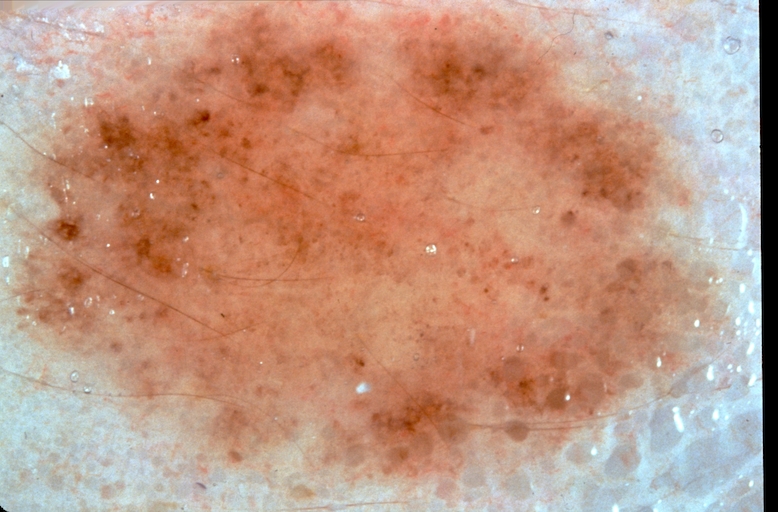}} &
		\subfloat[prediction for (a)]{\includegraphics[width = 1.0in]{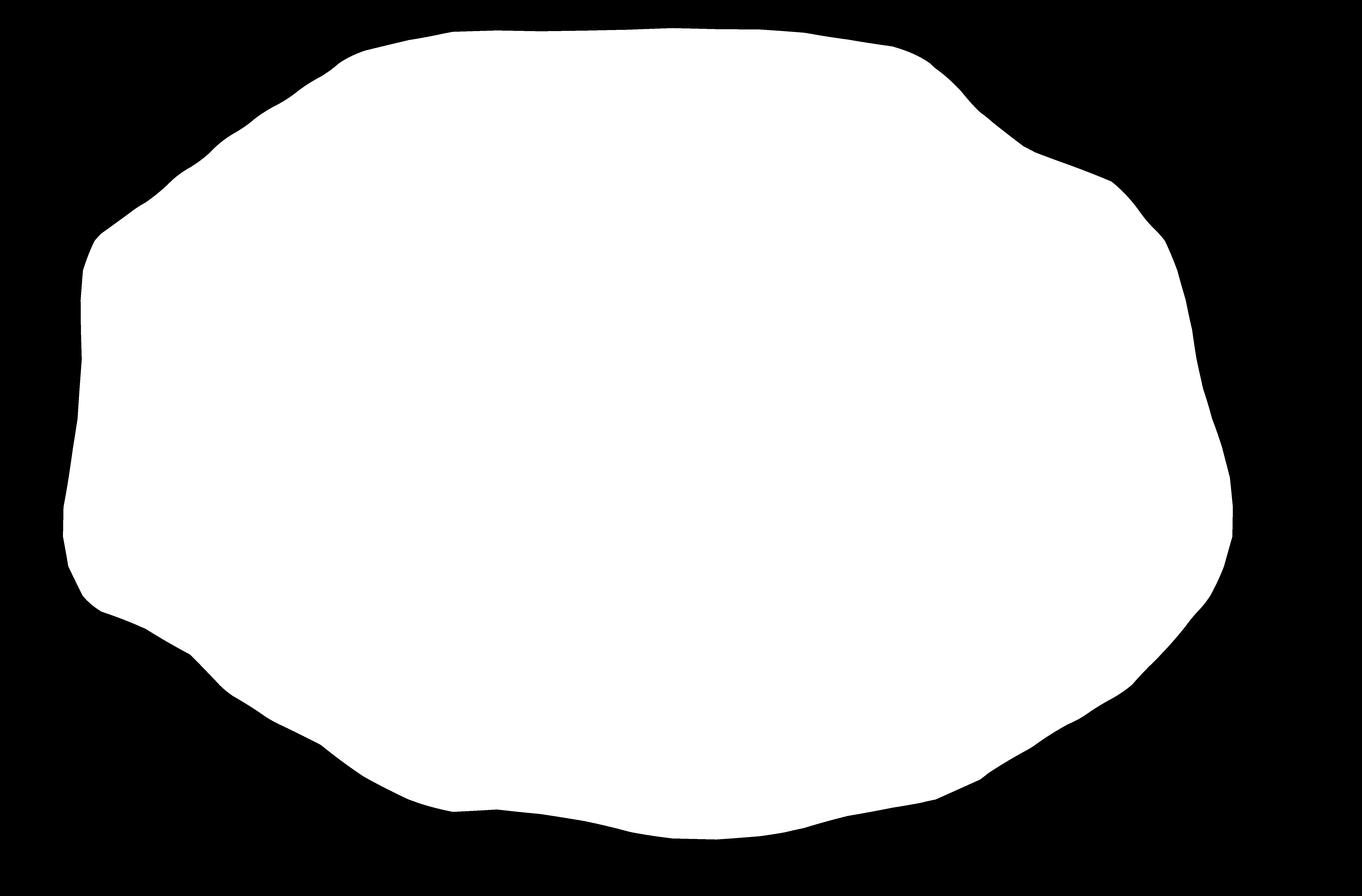}} &
		\subfloat[ground truth for (a)]{\includegraphics[width = 1.0in]{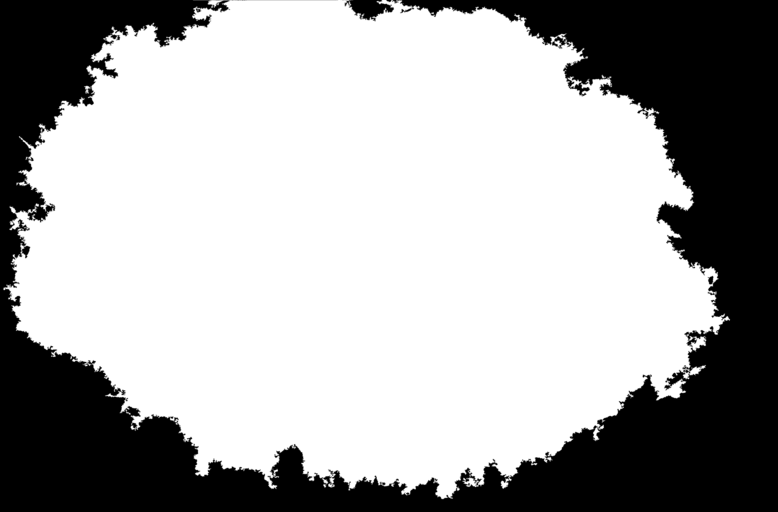}} \\
		\subfloat[input image]{\includegraphics[width = 1.0in]{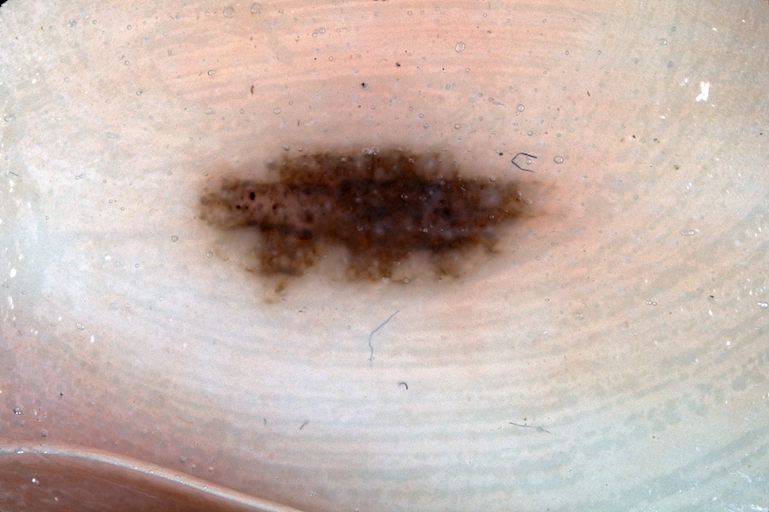}} &
		\subfloat[prediction (d)]{\includegraphics[width = 1.0in]{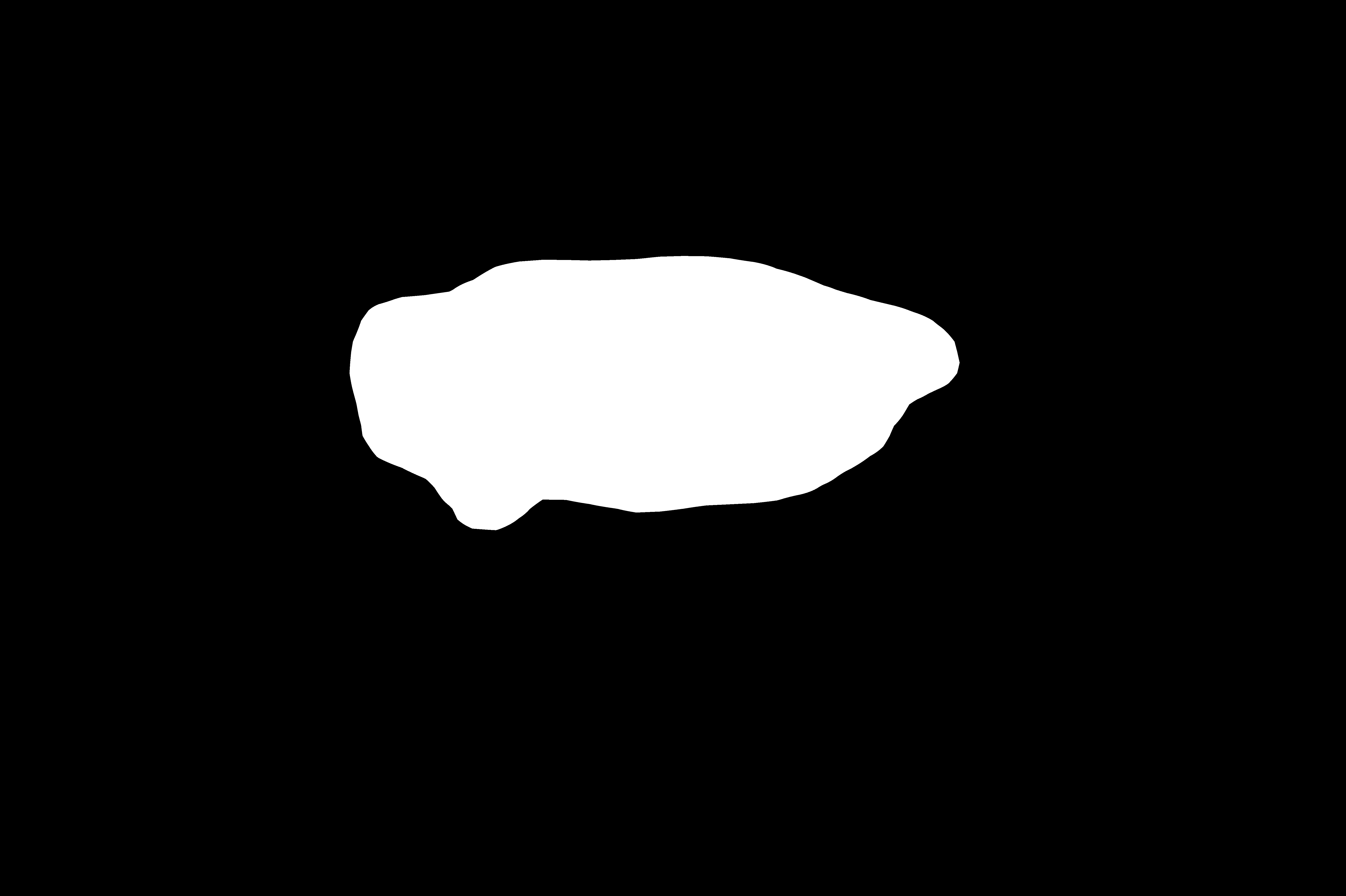}} &
		\subfloat[ground truth for (d)]{\includegraphics[width = 1.0in]{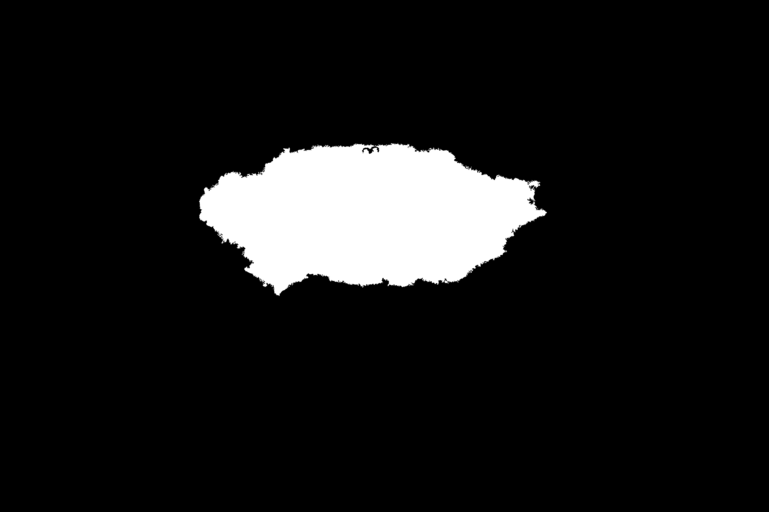}}\\
		\subfloat[input image]{\includegraphics[width = 1.0in]{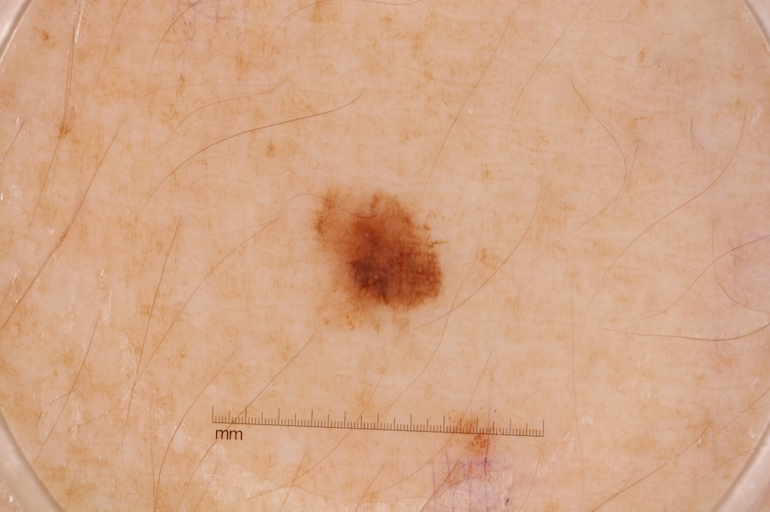}} &
		\subfloat[prediction for (g)]{\includegraphics[width = 	1.0in]{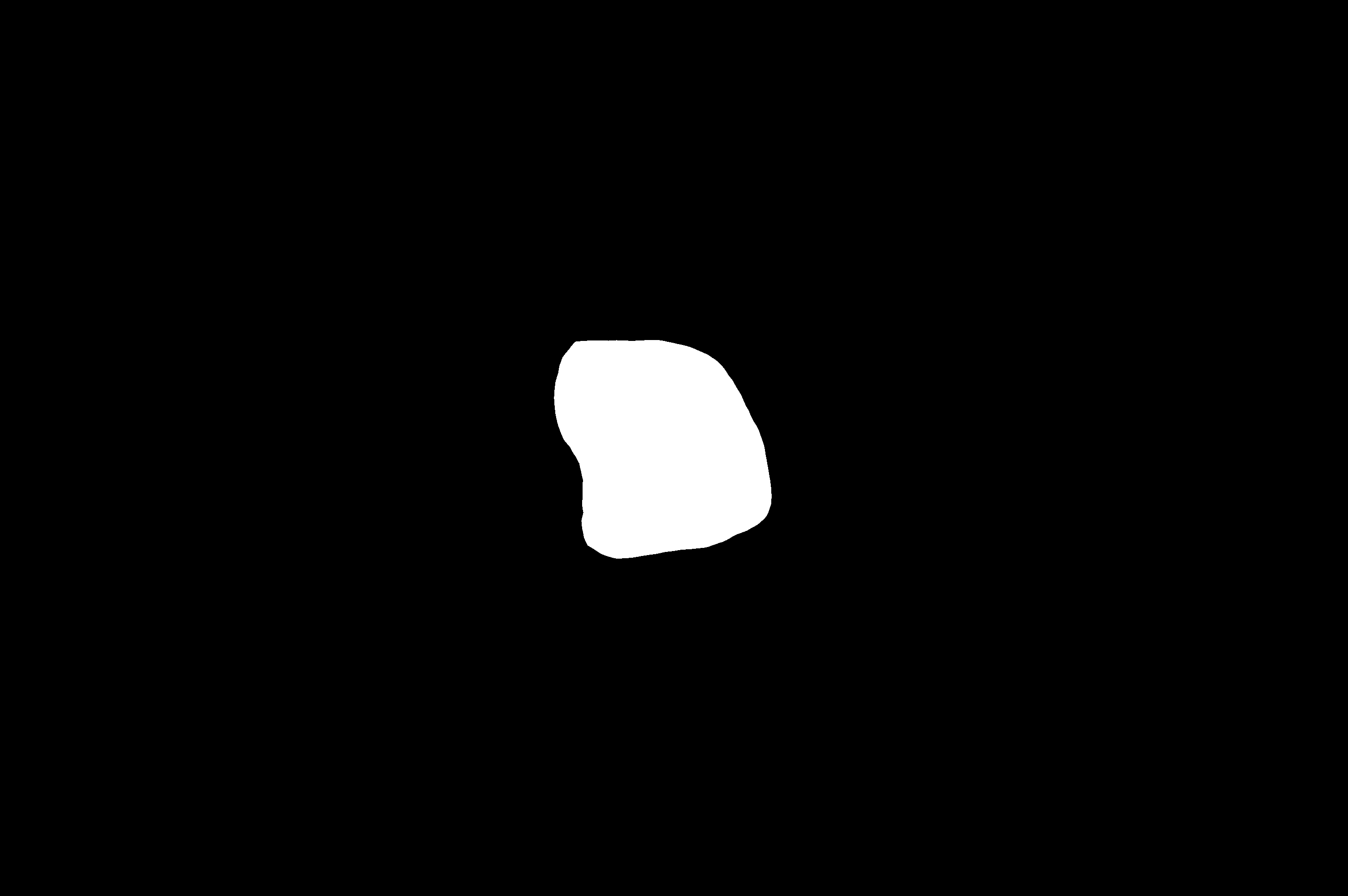}} &
		\subfloat[ground truth for (g)]{\includegraphics[width = 1.0in]{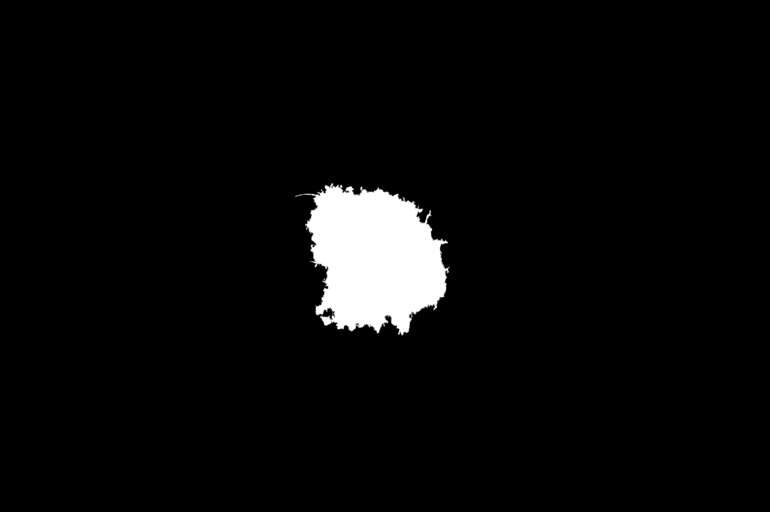}}\\	
	\end{tabular}
	\caption{Task 1 Lesion boundary detection results}
	\label{task1:results}
\end{figure}

Prior to the training, the mask R-CNN model was initialized using pretrained weights from COCO dataset\footnote{available at \url{https://github.com/matterport/Mask_RCNN/releases/download/v2.0/mask_rcnn_coco.h5}}\cite{DBLP:coco/journals/corr/LinMBHPRDZ14}.

In total, the training was performed for 40 epochs, the optimizer is a stochastic gradient descend algorithm with learning rate 0.001 and learning momentum 0.9.

\subsection{Test results}

In figure \ref{task1:results} a training image, a model prediction, and a ground truth is presented. Compared to the ground truth images the predictions capture less details due to the fixed mask shape -- 56 by 56 pixels.

The testing is performed on the images not included into the training dataset. In this work the score $ S_{1} $ of the model is calculated as an average over Jaccard indices of ground truth and predicted masks:

\begin{equation}
S_{1} = {{\sum\limits_{i=1}^N J(M_{g}(i), M_{p}(i))}\over{N}},
\label{eq:task1:score}
\end{equation}

\begin{flushleft}
	where $ N $ -- number of samples in test set, $ M_{g}(i) $ and $ M_{p}(i) $ - ground truth and predicted masks of i-th image correspondingly, $ J(\cdot, \cdot) $ -- Jaccard index (equation  \ref{eq:jccrd}).
\end{flushleft}

The model score $ S_{1} $ calculated for the training dataset is 0.7902. 

\subsection{Validation results}
 The model was also tested using validation data -- 100 images -- without ground truth, the overall score returned by submission system -- 0.789, which is coherent with the internal metrics (\ref{eq:task1:score}).

\section{Task 2: Lesion Attribute Detection}
\subsection{Configuration}

Compared to task 1 Mask-RCNN model configuration for task 2 differs only in the number of classes, which changed from 1 class indicating lesion boundary to 5 classes indicating lesion attributes.

Model training for task 2 was performed for 80 epochs.

\subsection{Test results}

In order to estimate the trained model for task 2 two scoring functions used in this project: $ S_{2}(j) $ -- returning averaged Jaccard index between ground truths and model predictions for class $ j $:

\begin{equation}
S_{2}(j) = {{\sum\limits_{i=1}^{N(j)} J(M^{j}_{g}(i), M^{j}_{p}(i))}\over{N(j)}},
\label{eq:task2:class_score}
\end{equation}

\begin{flushleft}
where $ 1 \le j \le 5 $ -- attribute class, $ N(j) $ -- number of non-empty ground truth masks for class j in test set, $ M^{j}_{g}(i) $ and $ M^{j}_{p}(i) $ -- ground truth and predicted masks for i-th image, indicating location of class $ j $ attributes, $ J(\cdot, \cdot) $ -- Jaccard index (equation \ref{eq:jccrd}).
\end{flushleft}

The overall model score used in this project is:

\begin{equation}
S_{2} = {1\over{5}}{\sum\limits_{j=1}^5 S_{2}(j)},
\label{eq:task2:score}
\end{equation}

The actual recognition results are unsatisfactory.

Model score  $ S_{2}(j) $ (\ref{eq:task2:class_score}) for:
\begin{enumerate}
	\item[$\bullet$] globules: 0.2610
	\item[$\bullet$] milia-like cysts: 0.2120
	\item[$\bullet$] negative network: 0.3082
	\item[$\bullet$] pigment network: 0.3725
	\item[$\bullet$] streaks: 0.2462
	\item[$\bullet$] average: 0.2800
\end{enumerate}

\begin{figure}[b!]
	\centering
	\begin{tabular}{cc}
		\subfloat[globules ground truth]{\includegraphics[width = 1.0in]{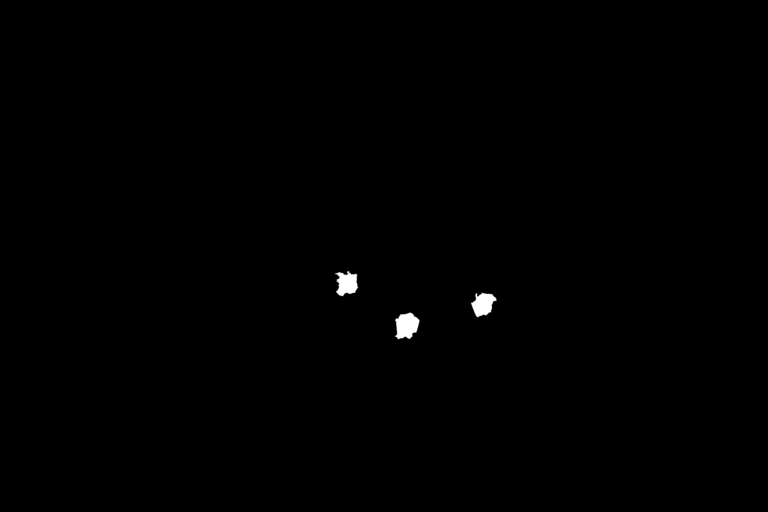}} &
		\subfloat[pigment network ground truth]{\includegraphics[width = 1.0in]{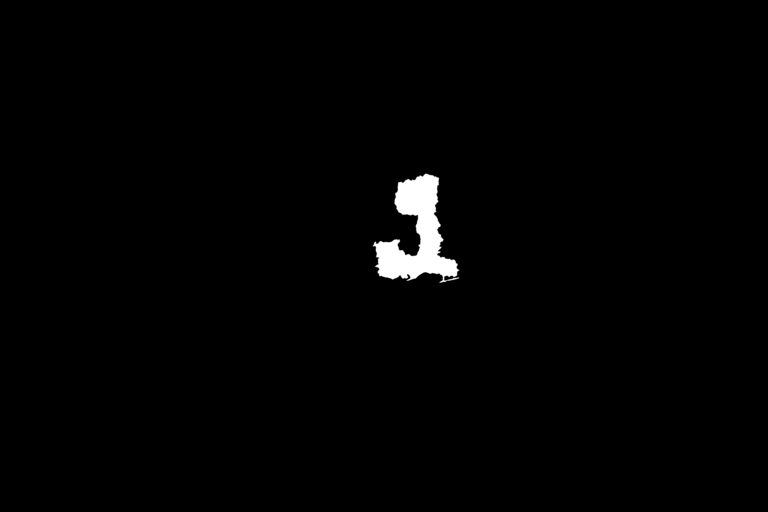}} \\
		\subfloat[globules prediction for (a)]{\includegraphics[width = 1.0in]{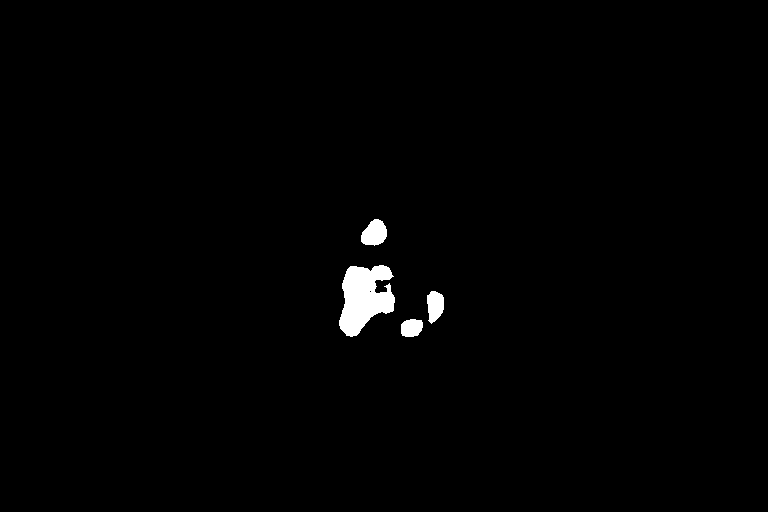}} &
		\subfloat[pigment network prediction for (b)]{\includegraphics[width = 1.0in]{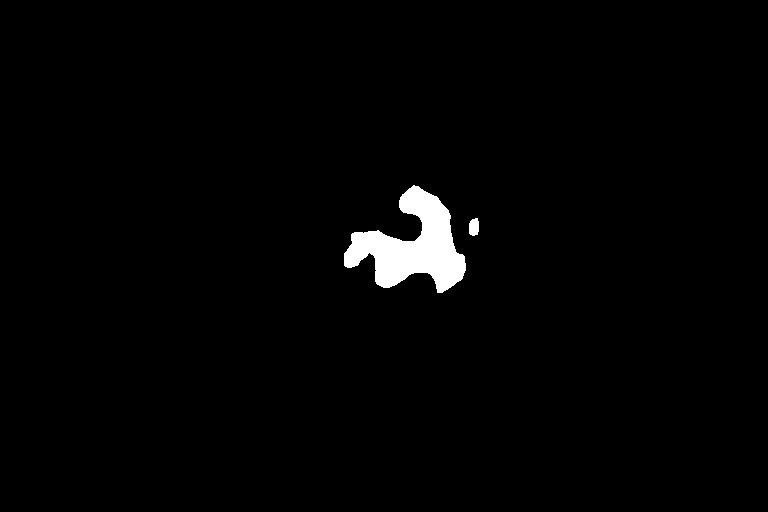}} \\
		\subfloat[globules ground truth]{\includegraphics[width = 1.0in]{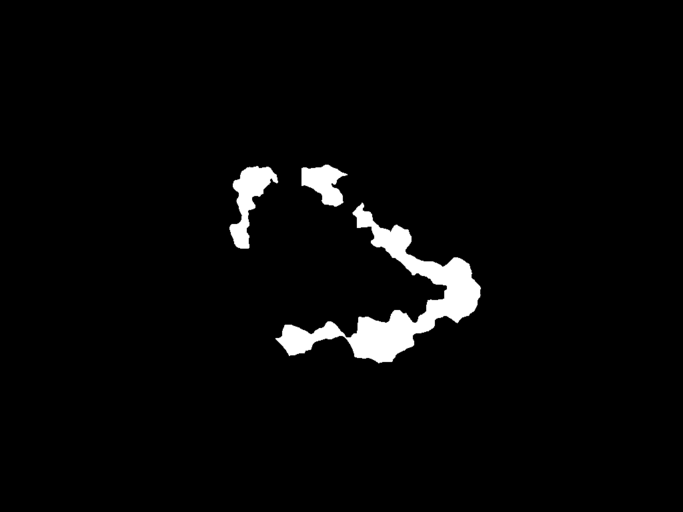}} &
		\subfloat[pigment network ground truth]{\includegraphics[width = 1.0in]{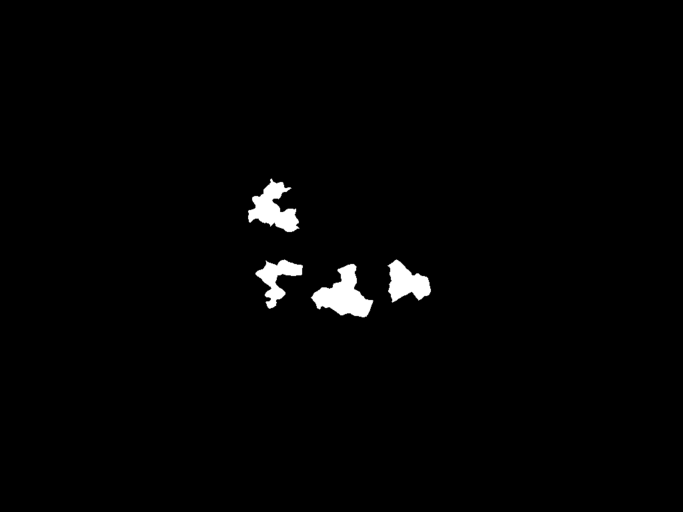}} \\		
		\subfloat[globules prediction for (e)]{\includegraphics[width = 1.0in]{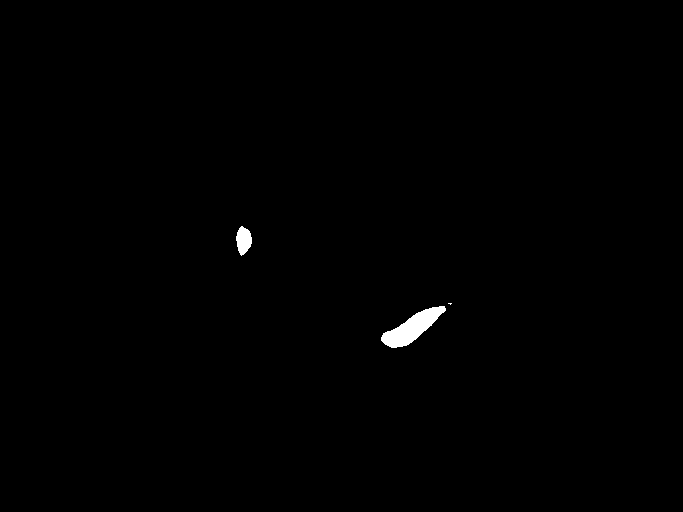}} &
		\subfloat[pigment network prediction for (f)]{\includegraphics[width = 1.0in]{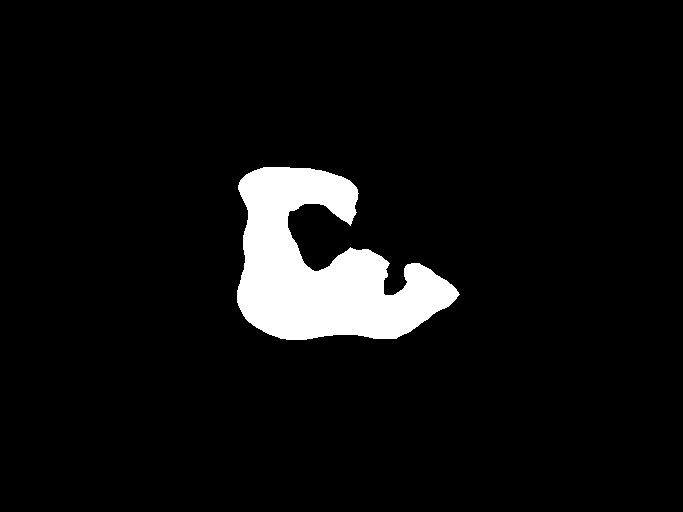}} \\				
	\end{tabular}
	\caption{Task 2 Lesion attribute detection errors}
	\label{task2:results}
\end{figure}

In figure \ref{task2:results} predicted and ground truth masks for an image are given. A visual analysis of similar images leads to a conclusion that Mask R-CNN method fails to detect boundaries accurately in case of complex overlays between different classes. It is also possible that the chosen resolution of the input layer is not high enough to capture, for instance, globules details and recognize smaller attributes of lesions.

\subsection{Validation results}
The model was also automatically tested using validation data -- 100 images -- without ground truth, the averaged validation score returned by submission system -- 0.409, which exceeds the internal score (\ref{eq:task2:class_score}) for every class prediction in the test set.

\section{Task 3: Disease Classification}
\subsection{Hybrid approach}

\begin{figure}[t!]
	\tikzstyle{block} = [rectangle, draw, fill=blue!20, 
	text width=5em, text centered, rounded corners, minimum height=4em, node distance=2cm]
	\tikzstyle{line} = [draw, -latex', line width = 0.1cm]
	\tikzstyle{cloud} = [draw, rectangle, node distance=2cm, minimum height=2em]
	
	\begin{tikzpicture}
	\node [cloud] (input) {\subfloat[input image]{\includegraphics[width=2.5 cm]{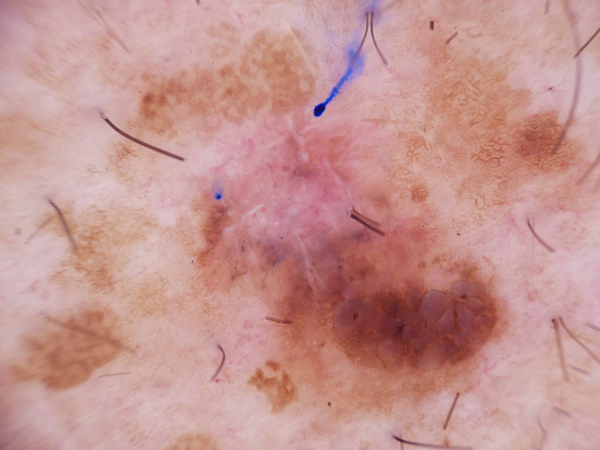}}};
	\node [block, right of = input, node distance=4cm] (model1) {Lesion boundary  detector};
	\node [cloud, below of=model1, node distance = 4cm] (genmask) {\subfloat[generated mask]{\label{task3:mask} \includegraphics[width=2.5cm]{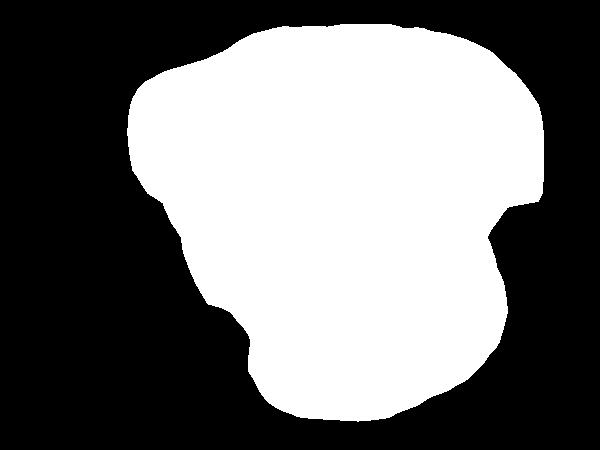}}};
	\node [block, below of=input, node distance = 4cm] (model3) {Lesion class mask detector};
	\node [cloud, below of=model3, node distance = 4 cm] (predclass) {\subfloat[class AKIEC predicted mask, area=0.3688]{\includegraphics[width=2.5 cm]{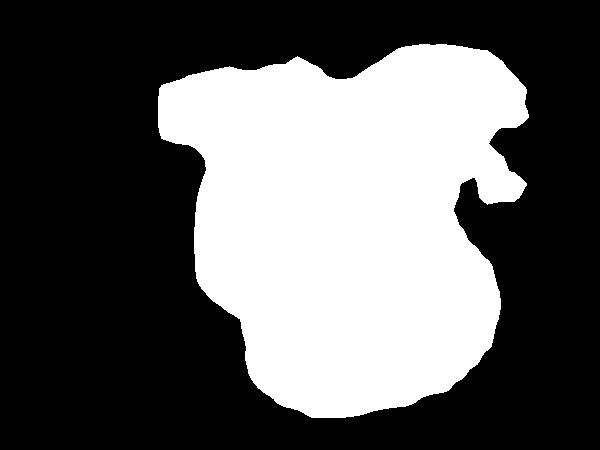}} \qquad \subfloat[class MEL predicted mask, area=0.3658]{\includegraphics[width=2.5 cm]{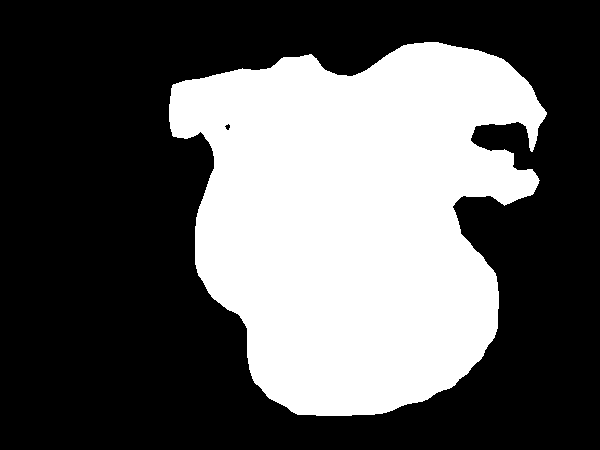}}};
	\node [cloud, below of=predclass, node distance = 4 cm] (final){final predicted class - AKIEC};
	\path [line] (input) -- (model1);
	\path [line] (model1) -- (genmask);
	\path [line] (genmask) -- (model3);
	\path [line] (input) -- (model3);
	\path [line] (model3) -- (predclass);
	\path [line] (predclass) -- (final);
	\end{tikzpicture}
	\caption{Task 3 Lesion class recognition process}
	\label{task3:hybrid}
\end{figure}
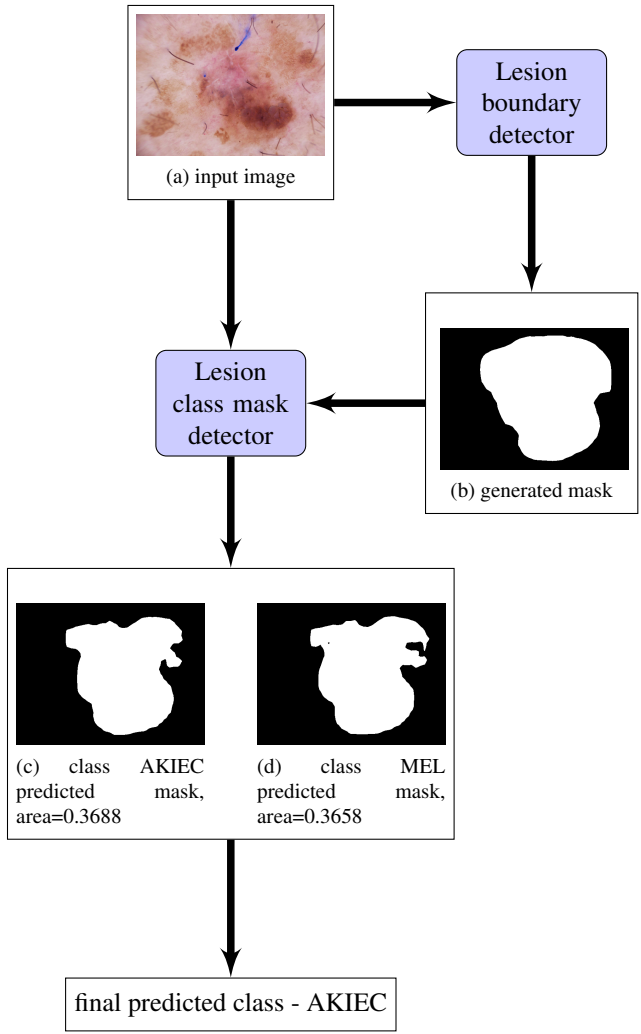

For this task, a set of lesion images and a CSV file describing disease type is provided for training.
The proposed solution to the disease classification problem is a hybrid approach described below (see figure \ref{task3:hybrid}).
\begin{enumerate}
	\item The model trained for lesion boundary detection (task 1) was used to build a lesion mask for every image in the training and validation set. (e.g. figure \ref{task3:mask})
	\item Every pixel of the masks from step 1 was assigned with a disease class contained in the ground truth CSV file.
	\item Lesion images and corresponding disease class masks were used as a training set for a model similar to the model described in a proposed solution to task 2, almost with the same parameters, except for the number of classes (7 instead of 5), and a different image resolution (task 3 input data dimensions are fixed at 600x450 pixels)
	\item The trained model from step 3 was used to classify disease:
	\begin{enumerate}
		\item model prediction for an input image includes 7 masks (one per each disease class)
		\item a mask with the biggest area of active pixels is chosen, the corresponding class of the selected mask is returned as model prediction
	\end{enumerate}
\end{enumerate}

\begin{figure}[b!]
	\resizebox{\columnwidth}{!}{%
		\begin{tabular}{l|l|l|l|l|l|l|l|l|}
			\multicolumn{2}{c}{} & \multicolumn{7}{c}{Predicted class}               \\		 
			\cline{3-9}
			\multicolumn{2}{c|}{}  &     MEL    & NV & BCC & AKIEC & BKL & DF & VASC \\ \cline{2-9}
			
			\multirow{7}{*}{\rotatebox[origin=c]{90}{Actual class}} & MEL   & 0.7544 & 0.1754 & 0.0132 & 0.0132 & 0.0395 & 0.0  &   0.0044   \\ \cline{2-9}
			& NV    &  0.0524 & 0.9248 & 0.0068 &  0.0015 &  0.0106 &  0.0023 &  0.0015  \\ \cline{2-9}
			& BCC   & 0.0693 & 0.0396 &  0.7228 & 0.0594  & 0.0594 & 0.0297 &  0.0198   \\ \cline{2-9}
			& AKIEC & 0.2308 & 0.0154 & 0.0769 & 0.4615 & 0.1538 & 0.0615 & 0.0      \\ \cline{2-9}
			& BKL   &  0.1595 & 0.125 &  0.0302 & 0.0086 & 0.6509 & 0.0259 & 0.0     \\ \cline{2-9}
			& DF    &   0.0714 & 0.25 &  0.0357 & 0.0714 & 0.0  &   0.5714  & 0.0      \\ \cline{2-9}
			& VASC  &  0.0 &     0.0667 & 0.0 &     0.0 &  0.0 & 0.0333 & 0.9      \\
			\cline{2-9}
	\end{tabular}}
	\caption{Task 3 Confusion matrix}
	\label{task3:cm}
\end{figure}

\subsection{Test results}

The score of the task 3 model $ S_{3}  $ is a ratio of correctly predicted diseases to a total number of training set images. On the training set provided,  $ S_{3} = 0.8420 $ . In figure \ref{task3:cm} a confusion matrix is given for the following diagnoses: 

\begin{enumerate}
	\item[$\bullet$] MEL -- Melanoma;
	\item[$\bullet$] NV -- Melanocytic nevus;
	\item[$\bullet$] BCC -- Basal cell carcinoma;
	\item[$\bullet$] AKIEC -- Actinic keratosis / Bowen’s disease (intraepithelial carcinoma);
	\item[$\bullet$] BKL -- Benign keratosis (solar lentigo / seborrheic keratosis / lichen planus-like keratosis);
	\item[$\bullet$] DF -- Dermatofibroma;
	\item[$\bullet$] VASC -- Vascular lesion.
\end{enumerate}

According to the confusion matrix, the best performance on the training set was achieved on NV class (Melanocytic nevus) detection with accuracy close to 92\%, the worst - for class AKIEC (Actinic keratosis / intraepithelial carcinoma) with 46\% accuracy of detection with 23\% of samples incorrectly classified as MEL (melanoma) class.

\subsection{Validation results}

The model was also automatically tested using validation data -- 193 images -- without ground truth, the averaged validation score returned by submission system -- 0.744, which is lower than  $ S_{3} $ score for the training set.

\bibliographystyle{ieeetr}
\bibliography{refs}

\end{document}